\documentclass{aa}
\usepackage[dvips]{graphicx}
\usepackage{txfonts}
\usepackage{amssymb}
\begin{document}
    \title{Exoplanet detection capability of the COROT space mission}

    \author{P. Bord\'e\inst{1}
          \and D. Rouan\inst{1}
          \and A. L\'eger\inst{2}
          }

    \offprints{P. Bord\'e}

    \institute{LESIA, UMR8109,  Observatoire de Paris, 5 place Jules Janssen,
              F-92195 Meudon \\
              \email{Pascal.Borde@obspm.fr, Daniel.Rouan@obspm.fr}
         \and
             Institut d'Astrophysique Spatiale, UMR8617, Universit\'e Paris XI,
             F-91405 Orsay \\
             \email{Alain.Leger@ias.fr}
             }

    \date{Received / Accepted}

\abstract{COROT will be the first high precision photometric
satellite to be launched with the aim of detecting exoplanets by
the transit method. In this paper, we present the simulations we
have carried out in order to assess the detection capability of
COROT. Using the model of stellar population synthesis of the
Galaxy developed at Besan\c con Observatory (Robin \& Cr\'ez\'e
\cite{robin}) and a simple cross-correlation technique (Bord\'e et
al. \cite{borde01}), we find that COROT has the capacity to detect
numerous exoplanets, not only Jupiter and Uranus-class ones, but
also hot terrestrial planets, if they exist. We show that small
exoplanets should be mainly gathered around 14--15th magnitude
K2--M2 dwarfs and giant exoplanets around 15--16th magnitude
F7--G2 dwarfs. We study the effect of crowding and the impact of a
high stellar variability noise that both reduce the detection
capability of the instrument.
\keywords{stars: planetary systems
-- methods: statistical -- techniques: photometric} }

\maketitle
%
%______________________________________________ Section 1 / Introduction
%
\section{Introduction}
To date about a hundred exoplanets have been discovered
spectroscopically by measuring the reflex motion of the star due
to the gravitational pull of its planet(s) (e.g. Perryman
\cite{perryman} for a review). The existence of one of these
planets was confirmed independently when the partial occultation
of the star by its planet was observed from the ground
(Charbonneau et al. \cite{charbonneau}), then from space (Brown et
al. \cite{brown}, Vidal-Madjar et al. \cite{vidal}). A number of
groups are looking for planets using various techniques from
the ground but owing to the disturbing effects of the turbulent
atmosphere, all exoplanets discovered so far are giant gaseous
planets whose masses are comparable to that of Jupiter or Saturn.
The main hope to collect a significant sample of Earth to
Uranus-class planets in the coming years is to go to space. This
is precisely one of the two main goals of the space mission COROT
to be launched in 2005.

In this paper, we are concerned with the instrumentation, data
analysis and expected performances of COROT in terms of exoplanet
detection. In Sect.~\ref{sec:mission}, we present the mission and
describe the characteristics of the instrumental set-up. In
Sect.~\ref{sec:method}, we review the detection method as it has
been implemented in our simulations in order to compute the number
of the expected detections. We discuss in Sect.~\ref{sec:det} the
detection efficiency of COROT as a function of the parent star
magnitude and spectral type, and in Sect.~\ref{sec:effects} some
effects that decrease the detection efficiency or may cause
spurious detections.
%
%_________________ Section 2 / The mission and its instrumental set-up
%
\section{The mission and its instrumental set-up} \label{sec:mission}
COROT was selected within the frame of the Small Mission Program
of the French space agency CNES. It will cost typically 63~Meuros.
Partners of CNES are several French laboratories: LAM (Marseille),
LESIA (Meudon), IAS (Orsay), and several European countries:
Austria, Spain, Belgium, and ESA/ESTEC. The goal of COROT is to
perform high accuracy photometry on a total field as wide as
7.0~deg$^2$ in order to fulfil the requirements of the two main
scientific objectives of the mission: a) measurement of stellar
pulsations on a limited set of stars (Baglin et al. \cite{baglin})
and b) detection of exoplanets transiting in front of their parent
star (Rouan et al. \cite{rouan}). COROT, the first transit mission
in space, will have at least two followers: Kepler (Koch et al.
\cite{koch}) on the American side, and Eddington (Favata et al.
\cite{favata}) on the European side.

The two programs share the same instrument, featuring a 27~cm
telescope without obscuration that includes two off-axis parabolas
and a dioptric objective. Detection is achieved thanks to four
cooled ($-40\degr$C) CCD 2048$\times$2048 from EEV, two devices
being dedicated to each program. The CCDs are arranged according
to an almost square pattern. The field of view of the exoplanets
program is 3.5~deg$^2$. The orbit of COROT is pseudo-polar,
quasi-inertial, at an altitude of $\approx 900$~km. The generic
platform, PROTEUS, provides a pointing stability of $\approx
0.2\arcsec$ when the instrumental information on the star
positions is used. During the 2.5 years mission, five fields will
be observed continuously, each one during 150 days. In order to
prevent the limb of the Earth from being a source of background,
COROT will be pointing in a direction not far from the equatorial
plane. Since the Sun should also be avoided, the five fields are
grouped around two main directions: one close to the Galactic
Center ($l_\mathrm{II}=35\degr$, $b_\mathrm{II}=0\degr$) and one
close to the Anticenter ($l_\mathrm{II}=210\degr$,
$b_\mathrm{II}=0\degr$). After six months of observation in one of
those directions, the satellite is rotated by 180$\degr$ with
respect to the polar axis. The five fields will therefore be
either 3 in the Center direction and 2 in the Anticenter direction or
the reverse. The decision will be taken according to the launching
date.

The goal of the exoplanet program will be achieved by monitoring
continuously up to 12000 dwarf stars in each field, with visual
magnitudes from V=11 to V=16. The fields have been chosen at
rather low galactic latitudes (b$_\mathrm{II}<10\degr$) in order
to have a large density of stars. The technique used is aperture
photometry: the flux collected during an elementary exposure of
32~s is measured by summing all pixels in a fixed mask
encompassing the star Point Spread Function (PSF). The PSF
covers 100--105~px at V=11--12, 80--90~px at V=13--14, 40--50~px
at $V \ge 15$, and is itself a very low resolution on-axis
spectrum of the star formed by a small bi-prism inserted a few
centimeters above the exoplanet CCDs. For $V \le 14$, this
additional device provides color information in three bands
(three subsets of the PSF pixels) that will be used as a powerful
diagnostic tool to analyze doubtful events and may help to remove
some of the stellar variability noise.

In order to cope with the data transmission rate of 1500~Mbits per
day, data will be co-added on-board during periods of 8.5 minutes
(16 exposures), before being downloaded. The integration will also
be synchronized on the orbital period, so that any perturbation at
the orbital frequency, such as the thermal fluctuations generated
by Sun eclipses along the orbit, could be cancelled out at first
order by summing the whole set of data taken during one orbital
period. Time sampling will be improved, down to 32~s (one
exposure) for any star where a high signal-to-noise event, like
the transit of a giant planet, will be detected. This will make it possible
to measure short-duration effects, such as limb-darkening on the
stellar disk or variations of the transit period due to satellites
around the planet or to other (non-occulting) planets (Sartoretti
\& Schneider \cite{sartoretti}). Scattered light from the Earth
limb is maintained at a low level thanks to the afocal telescope
design and to a long baffle. The largest contribution among all
sources of background originates from the zodiacal light that was
evaluated to be no more than
${16\;\mathrm{e}^-\;\mathrm{s}^{-1}~\,\mathrm{px}^{-1}}$, using
data from James et al. (\cite{james}).
%
%__________________________________________ Section 3 / Detection method
%
\section{Detection method}  \label{sec:method}
Exoplanet detection by the way of transit observations has been
described many times in the literature (Rosenblatt
\cite{rosenblatt}, Borucki \& Summers \cite{borucki84}, Jenkins et
al. \cite{jenkins96}). Let us remind the reader that at least
three identical and equally spaced dimmings in a stellar light
curve are interpreted as the signature of an occulting planetary
companion. As the dimming relative depth is in the ratio of the
surface of the planet to that of the star, it gives a measure of
the planetary radius. In a previous paper (Bord\'e et al.
\cite{borde01}) we have used a simple cross-correlation treatment,
equivalent to the matched filter approach of Jenkins et al.
(\cite{jenkins96}), to obtain a preliminary evaluation of the
mission performances. Here, we will only review the key points of
this method.

For the purpose of pure detection, especially at low S/N
(signal-to-noise ratio), the transit signal can be described by 4
parameters: the amplitude $A$ and duration $\tau$ of an individual
transit, the transit period $P$ (equal to the revolution period of
the planet), and the phase of the first transit $\phi$ (i.e. its
date expressed as a fraction of $P$). The following method
explores the $(\tau,P,\phi)$ parameter space by evaluating the
likelihood of every triplet and gives the amplitude $A$ as a
by-product. A tested triplet is referred to as a trial triplet.

For COROT targets, $\tau$ will be of the order of a few hours,
certainly remaining below ${\tau_\mathrm{max}=15\;\mathrm{h}}$.
Thus, the first step in the data processing will be to high-pass
filter the light curves with a cut-off frequency of say
$1/(4\,\tau_\mathrm{max})$, in order to remove the irrelevant long-term
stellar variations.  The light curves are then averaged on a
trial transit duration $\tau$ to increase the S/N, and
cross-correlated with a transit-like signal at a trial period $P$.
This noise-free signal has the shape of a comb with $k$ teeth, $k$
being the number of transits with a period $P$ during a 150-day
exposure on a given stellar field. Cross-correlation products
$C(\tau,P,\phi)$ have to be computed for enough trial triplets to
correctly explore the parameter space. Detection occurs if a
threshold fixed by a given confidence level is exceeded: $C \ge
\beta \, \sigma_C$, where $\sigma_C$ is the standard deviation of
the noise affecting $C$. We find that our requirement of less than
one false detection for the entire mission is met for ${\beta =
7}$, if a Gaussian statistics is assumed for $C$. This value is
conservative as our parameter grid was not built to avoid
correlated values of $C$. Indeed, our goal here is not to have a
refined data processing as discussed recently by Jenkins et al.
(\cite{jenkins02}) but instead to reach a first order estimate of
the detection capability of COROT.

The minimum S/N on a \emph{single transit} necessary to assess a
detection can be deduced from the above cross-correlation
criterion:
\begin{equation} \label{eq:snr}
\frac{S}{N} \ge \frac{\beta}{\sqrt{k}}.
\end{equation}
In turn, Eq.~(\ref{eq:snr}) leads to the minimum radius the planet
should have to be detected:
\begin{equation} \label{eq:Rp}
R_\mathrm{P} \ge R_\star {\left( \frac{\beta}{\sqrt{k}}
\frac{\sigma_n}{N_e} \right)}^\frac{1}{2},
\end{equation}
where $R_\star$ is the radius of the star, $\sigma_n$ the standard
deviation of the noise affecting the light curve and $N_e$ the
number of photo-electrons collected during $\tau$. In our
simulations, we include the quantum noise, the read-out noise, the
background noise (zodiacal light), the jitter noise and the
stellar variability. The jitter noise, i.e. the
photometric variations caused by the wobble of the PSF with
respect to its photometric mask due to satellite pointing errors,
will be corrected afterwards using the data from the seismology
field. It amounts to $\approx 40 \%$ of the quantum noise. The
stellar variability is assumed to be identical to that of the Sun
and estimated to be ${\simeq 50 \; \mathrm{ppm}}$ rms at our
timescale of interest using the SOHO data (Fr\"ohlich et al.
\cite{frohlich}).
%
%_______________________________________ Section 4 / Expected detections
%
\section{Expected detections}   \label{sec:det}
\subsection{Synthetic stellar field}
\begin{figure}
\centering
\includegraphics[width=8cm]{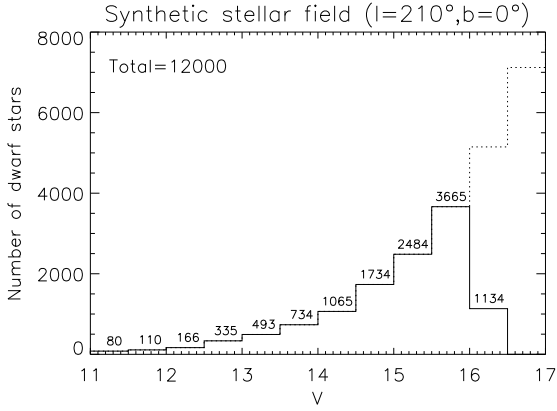}
\caption{Distribution of dwarfs in the 3.5~deg$^2$
synthesized field vs. the visual magnitude. Dashed line: complete
Besan\c con model (16000 stars for $V \le 16.5$). Solid line:
selection of the 12000 brightest targets for COROT.}
\label{fig:bes_V}
\end{figure}

\begin{figure}
\centering
\includegraphics[width=8cm]{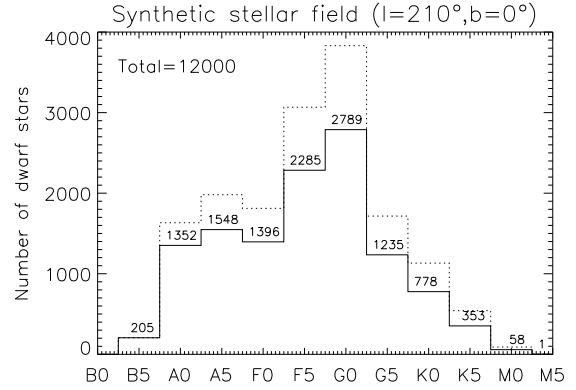}
\caption{Same as Fig.~\ref{fig:bes_V} vs. the spectral
type.} \label{fig:bes_Sp}
\end{figure}

In order to assess the performance of COROT, we need realistic
distributions of stars as a function of their visual magnitude $V$
and their spectral type Sp in the directions monitored by the
satellite. Such distributions were synthesized with the stellar
model of the Milky Way developed by Robin \& Cr\'ez\'e
(\cite{robin}, regularly updated), and retrieved from Besan\c con
Observatory through their web interface. The 3.5~deg$^2$ target
field for the simulations in this paper is centered in the
direction of the Galactic Anticenter ($l_\mathrm{II}=210\degr$,
$b_\mathrm{II}=0\degr$). It contains $\approx 16000$ dwarfs with
magnitudes between 11 and 16.5, and spectral types between B0 and
M5. The resolution used is 0.5 in magnitude, e.g. ${V=14}$
corresponds to ${13.75 \le V \le 14.25}$, and every
spectral type is divided into 4 subclasses, e.g. G1 corresponds to
G0~$\le$~Sp~$\le$~G2, G3 to G2~$\le$~Sp~$\le$~G4, and so on. The
number of dwarfs as a function of $V$ roughly follows a geometric
progression with a ratio of 2.0--2.5 (Fig.~\ref{fig:bes_V}) and
the whole sample peaks at spectral types F5--G0
(Fig.~\ref{fig:bes_Sp}). Owing to the capacities of
on-board electronics and telemetry of COROT, the total number of
targets will be limited to $\approx 12000$ stars.

In this paper, we present only simulations regarding the Galactic
Anticenter for which the Besan\c con model (with $A_V =
0.7$~mag/kpc) seems to agree well with preliminary stellar
counts in actual fields. Thus, for the purpose of this work, we
assume that COROT will observe 5 fields with statistical
properties identical to those of this field. In the
remainder of the paper, all figures regarding detection numbers
are given for the entire mission: 5 fields, 150 days each, 60000
dwarfs in total.

Although stars of other luminosity classes (giants, subgiants...)
are also present in the actual fields, they are not taken into account for
the prospect of planet detection, since their large radii would
lead to too weak transit signals. However, they contribute to the
crowding effect and may induce an additional variability discussed
in Sect.~\ref{sub:crowding}. When dealing with the observed
fields, stars will be selected according to their luminosity
classes, thanks to either specific ground observations (BVRI
photometry) or the use of DENIS and 2MASS catalogues. For example,
the DENIS survey (e.g. Epchtein et al. \cite{epchtein99}) will
provide the infrared photometry in the I, J and K bands down to
magnitudes I=18.5, J=16.5 and K=13.5. Most giant stars in
COROT fields could then be identified through the computations of
their infrared colors (Epchtein et al. \cite{epchtein97}), the
others by way of dedicated observations if necessary. The
COROT entry catalogue is currently being implemented by the
Laboratoire d'Astrophysique de Marseille (LAM).
\subsection{Detections at a given orbital distance}
Since at least three transits are needed, we can compute the
maximum distance at which a detection can occur as a function of
the spectral type of the parent star. In the prospect of
exobiology that is concerned with habitable zones around stars
(e.g. Kasting et al. \cite{kasting}, Franck et al.
\cite{franck}), let us introduce the reduced orbital radius
defined by ${a_\mathrm{r} = a {(L_{\star}/L_{\odot})}^{-0.5}}$, so
that a planet at ${a_\mathrm{r}=1}$~AU would receive as much flux
from its star as the Earth from the Sun. Equivalently, one can use
its effective blackbody temperature ${T_\mathrm{P} (\mathrm{K}) =
278 / \sqrt{a_\mathrm{r}}}$ (equal to the ground temperature only
if the albedo is zero and without any greenhouse effect). As for a
150 day observation the revolution period $P$ should be less than
75 days (maximum value for 3 transits and ${\phi=0}$), we draw the
accessible range of reduced orbital radii reported in
Table~\ref{tab:armax}. Based on the simulations by Franck
et al. (\cite{franck}), we conclude that depending on the age of
the star, COROT could detect planets in the habitable zone around
late K to early M dwarfs, if such planets exist.
\begin{table}[htbp]
\centering
\begin{tabular}{ccc}
\hline
\hline
Sp  & $a_\mathrm{r}^\mathrm{max}$ (AU) & $T_\mathrm{P}^\mathrm{min}$ (K) \\
\hline
A1V & 0.08 & 970 \\
F1V & 0.18 & 650 \\
G1V & 0.32 & 490 \\
G6V & 0.41 & 430 \\
K1V & 0.57 & 370 \\
K6V & 0.78 & 315 \\
K8V & 0.90 & 290 \\
M1V & 1.21 & 250 \\
\hline
\end{tabular}
\caption{Accessible range of reduced orbital distance/planet
effective temperature as a function of the spectral type of the
parent star.} \label{tab:armax}
\end{table}

\begin{figure}
\centering
\includegraphics[width=8cm]{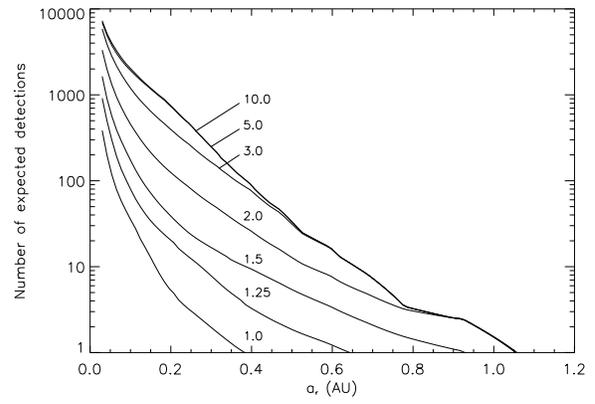}
\caption{Number of expected detections for the entire mission as a
function of the reduced orbital distance for various planetary
radii (expressed in unit of the Earth radius). It is assumed that
every star has one planet of the labelled radius positioned at the
considered distance.} \label{fig:det_ar}
\end{figure}

\begin{figure}
\centering
\includegraphics[width=8cm]{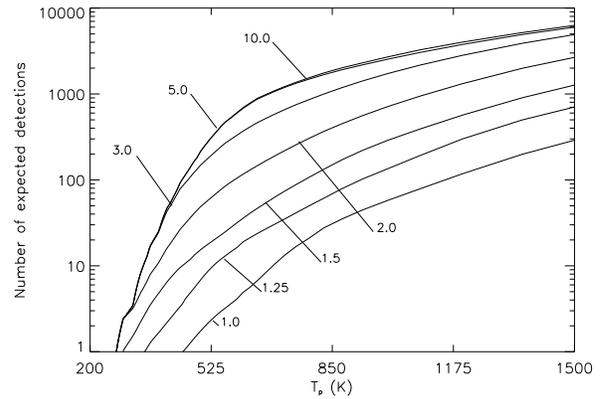}
\caption{Number of expected detections for the entire mission as a
function of the planet effective temperature for various planetary
radii (expressed in unit of the Earth radius). It is assumed that
every star has one planet of the labelled radius with the
considered effective temperature.} \label{fig:det_Tp}
\end{figure}

Now, let us assume that every dwarf in the synthetic field is
orbited by one planet. The number of detected planets by COROT
depends on the radii and (reduced) orbital distances of these
planets, of the magnitude and spectral type of their parent stars
and of the probability that at least three transits can be
observed during a 150-day run. As a first hypothesis, we assume
that all planets have the same radius $R_\mathrm{P}$ and are
positioned at the same reduced orbital distance $a_\mathrm{r}$.
Then, the number of detections reads $n_\mathrm{det} = n_\star
\times p_\mathrm{g} \times p_3$, where $n_\star$ is the number of
stars for which the criterion given by Eq.~(\ref{eq:Rp}) is met,
$p_\mathrm{g}$ the geometrical probability that transits are
observable and $p_3$ the probability to observe at least 3 of
them. Figures~\ref{fig:det_ar} and \ref{fig:det_Tp} display
detection curves, parameterized by the planet radius, as a
function of $a_\mathrm{r}$ or $T_\mathrm{P}$. Selected results are
also reported in Table~\ref{tab:det_ar}. The ripples appearing for
high $a_\mathrm{r}$ on Fig.~\ref{fig:det_ar} or low $T_\mathrm{P}$
on Fig.~\ref{fig:det_Tp} are an effect of the finite size of the
spectral type bins, and of the requirement on a minimum of
3 transits.
\begin{table}[htbp]
\centering
\begin{tabular}{ccrrrrr}
\hline
\hline
$a_\mathrm{r}$ (AU) & $T_\mathrm{P}$ (K) & 1~R$_\oplus$ &
1.5~R$_\oplus$ & 2~R$_\oplus$ & 3~R$_\oplus$ & 5~R$_\oplus$ \\
\hline
0.05 & 1200 & 120 & 570 & 1320 & 2800 & 3800 \\
0.14 &  750 &  17 &  90 &  260 &  750 & 1300 \\
0.30 &  500 &   2 &  17 &   55 &  160 &  240 \\
0.86 &  300 &   0 &   1 &    3 &    3 &    3 \\
1.00 &  278 &   0 &   1 &    2 &    2 &    2 \\
\hline
\end{tabular}
\caption{Selection of expected detections for various planetary
radii and reduced distances to the parent star/planet effective
temperatures, assuming every star has one planet for such
$a_\mathrm{r}$ and $R_\mathrm{P}$. Figures are given for the
entire mission and a confidence of less than one false detection.
The crowding effect (Sect.~\ref{sub:crowding}) is expected to
remove $\approx 10\%$ from the detections quoted here.}
\label{tab:det_ar}
\end{table}

It appears from these computations that terrestrial planets
(1--2~R$_\oplus$) are within reach of COROT provided their
effective temperatures are $\gtrsim 500$~K. As can be expected
from equation (\ref{eq:snr}), the closer the planet, the higher
the number $k$ of transits and the amount of detections. Besides,
we can tell from the almost superimposed 5 and 10~R$_\oplus$
curves that COROT reaches its full discovery potential as soon as
${R_\mathrm{P} \gtrsim 5 \, \mathrm{R}_\oplus}$.
\subsection{Integrated number of detections}  \label{sub:laws}
As a second hypothesis, we estimate the expected
detections by integrating along the $a_\mathrm{r}$ coordinate.
This can be done provided an orbital distribution of planets
around their parent stars is assumed. We have considered
\textit{a uniform orbital distribution law}: the planet
probability density as a function of $a_\mathrm{r}$ is constant
beyond $a_\mathrm{r} = a_\mathrm{r}^\mathrm{min}$ and
\textit{normalized to one planet per star per AU} (arbitrary). To
date, the actual observed value for $a_\mathrm{r}^\mathrm{min}$ is
0.03~AU (Schneider \cite{schneider}). This is consistent with the
simulations of planet migration by Trilling et al.
(\cite{trilling}). As the detection efficiency increases rapidly
with the proximity of the planet to its star
(Fig.~\ref{fig:det_ar}), the integrated number of detections is
very sensitive to $a_\mathrm{r}^\mathrm{min}$. In
Table~\ref{tab:det_int}, we give the results for
$a_\mathrm{r}^\mathrm{min} = 0.03$ and 0.05~AU.
\begin{table}[htbp]
\begin{center}
\begin{tabular}{rr}
\hline
\hline
$R_\mathrm{P}$  &  Integrated number \\
(in R$_\oplus$) &  of detections \\
\hline
  1.0   &    5--6    \\
  1.25  &   12--18   \\
  1.5   &   26--37   \\
  2.0   &   70--95   \\
  3.0   &  189--240  \\
  5.0   &  300--367  \\
(10.0)  & (311--382) \\
\hline
\end{tabular}
\end{center}
\caption{Integrated number of detections for the entire
mission as a function of the planetary radius (the confidence
level corresponds to less than one false detection). The higher
value corresponds to $a_\mathrm{r}^\mathrm{min} = 0.03$~AU and the
lower to $a_\mathrm{r}^\mathrm{min} = 0.05$~AU. The crowding
effect (Sect.~\ref{sub:crowding}) is expected to remove $\approx
10\%$ from the detections quoted here. The value for 10~R$_\oplus$
is given for completeness as the frequency of such planets is
known to be much less than 1 planet per star per reduced AU.}
\label{tab:det_int}
\end{table}
At this stage, it is very important to recall that these figures
hold for one planet per star. For giant planets around G stars,
radial velocity observations have shown that this assumption
largely overestimates the actual number of planets.
Considering that for giant planets (${R_\mathrm{P} \simeq
10 \, \mathrm{R}_\oplus}$) the radial velocity (RV) detections are
complete in the range 0--0.1~AU, $\approx 18$ planets (Schneider
\cite{schneider}) have been detected out of $\approx 2600$ stars
(Table~1 in Tabachnik \& Tremaine \cite{tabachnik}). For a uniform
orbital distribution, this translates into a frequency of $\approx
7\%$ in the range 0--1~AU and leads for COROT to $\approx$~25
detections. However, because of the high level of confidence that
is chosen, these figures should constitute the minimum that can be
expected from COROT. Inversely, \emph{the comparison between these
predictions and the actual number of planets detected by COROT
will inform us about the frequency of these planets}.
\subsection{Spectral type and magnitude of the parent star}
\begin{figure*}
\centering
\includegraphics[width=12cm]{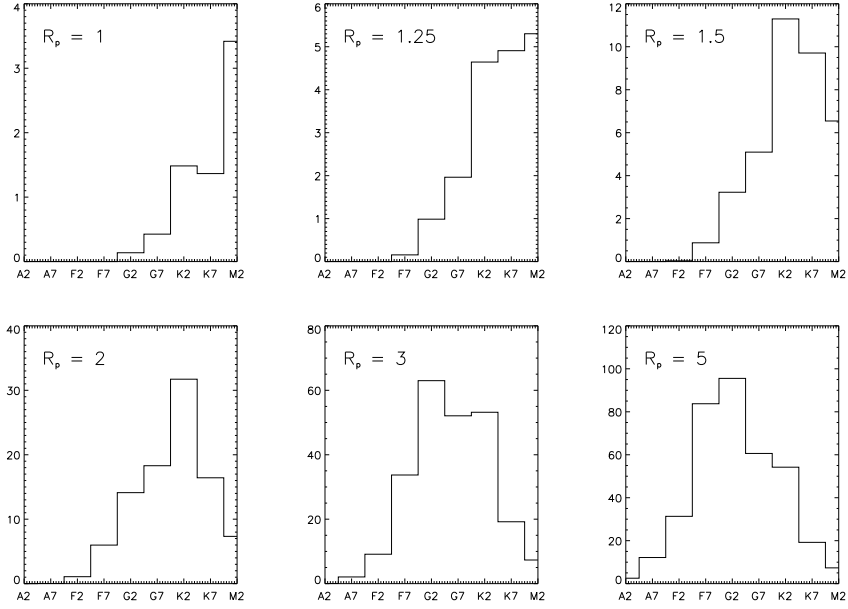}
\caption{Histograms of expected detections for the whole
mission -- with a  false alarm rate less than one false detection
-- vs. the parent star spectral type, for various planetary radii
(expressed in unit of the Earth radius). The planets are assumed
to be uniformly distributed as a function of their orbital
distances. The crowding effect (Sect.~\ref{sub:crowding}) is
expected to remove $\approx 10\%$ from the detections plotted
here.} \label{fig:det_sp}
\end{figure*}

\begin{figure*}
\centering
\includegraphics[width=12cm]{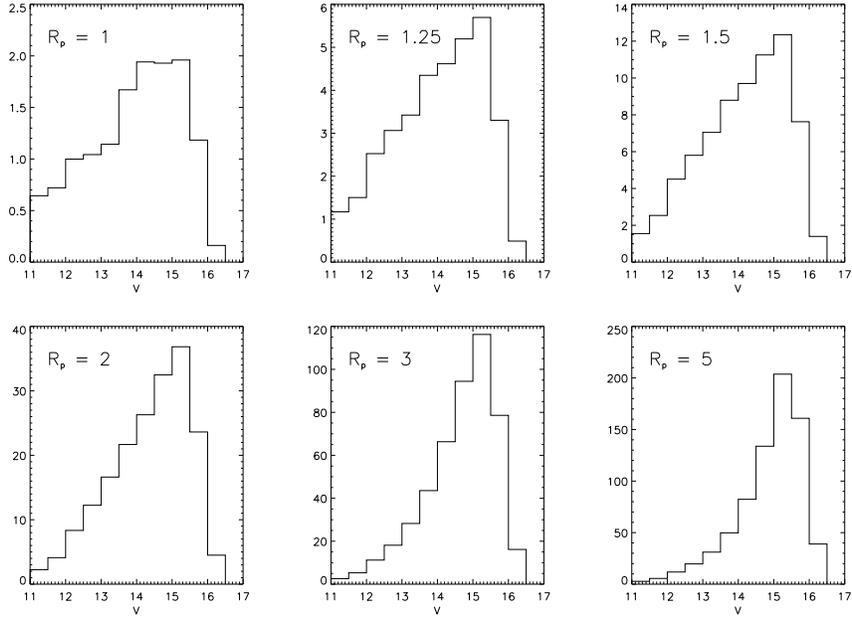}
\caption{Same as Fig.~\ref{fig:det_sp} vs. the visual magnitude of
the parent star.} \label{fig:det_mv}
\end{figure*}

Up to this point, all the results drawn from our stellar sample
were integrated on the spectral type Sp and the magnitude $V$ of
the parent star. Our concern is now to analyze the influence of
these two parameters on the detection capability of COROT. We have
computed histograms of the number of detections vs. Sp and $V$ for
different planetary radii
(Figs.~\ref{fig:det_sp}--\ref{fig:det_mv}). Histograms for
${R_\mathrm{P} = 10 \, \mathrm{R}_\oplus}$ are not significantly
different from those for ${R_\mathrm{P} = 5 \, \mathrm{R}_\oplus}$
and are not reproduced here.
With respect to the spectral type, we note that the detection peak
shifts progressively from M2 to K2 for terrestrial planets
(${R_\mathrm{P} \lesssim 2 \, \mathrm{R}_\oplus}$), then to G2 for
Uranus-class objects (${R_\mathrm{P} \simeq 3 \,
\mathrm{R}_\oplus}$) and F7--G2 for giant planets (${R_\mathrm{P}
\gtrsim 3 \, \mathrm{R}_\oplus}$). This last feature can be
attributed to the properties of the stellar sample itself that
peaks around F5--G0. The histograms vs. $V$ show a general
shape that strongly reflects that of the original stellar
distribution (Fig.~\ref{fig:bes_V}). As one goes deeper in
magnitude, the number of detections keeps increasing because on
one hand there are more stars, and on the other hand the detected
planets are essentially hot objects frequently transiting in front
of their parent stars. Therefore, we conclude that, for
pure detection purposes, COROT targets should primarily be chosen
among dwarfs later than F2 with magnitudes up to 16.5. This would
lead to a slight improvement of the numbers given in
Table~\ref{tab:det_int}.
%
%_______________ Section 5 / Effects decreasing the detection efficiency
%
\section{Effects decreasing the detection efficiency} \label{sec:effects}
\subsection{Stellar variability}
It is recognized that compared to other dwarfs, the Sun is a
pretty quiet star. Since the Sun has been chosen as the prototype
for our model of stellar variability, one may wonder what would
happen if this noise source were moderately or even considerably
higher. We have investigated the consequences of such an increased
variability on the integrated number of detections
(Sect.~\ref{sub:laws}) by computing the number of detections for a
stellar variability 10 and 50 times higher than the solar
value (Table~\ref{tab:var}). It is found that a high variability
(${\times 10}$) causes a loss of terrestrial planets by a factor
3--4, whereas a very high variability (${\times 50}$) makes their
detection out of reach. In that case however, we consider taking
advantage of the color information contained in the PSF
(Sect.\ref{sec:mission}) to remove part of this noise with a
proper combination of the colored channels (Bord\'e et al.
\cite{borde03}).
\begin{table}[htbp]
\begin{center}
\begin{tabular}{rrrr}
\hline
\hline
$R_\mathrm{P}$  &  solar var.  &  solar var.  & solar var.   \\
(in R$_\oplus$) &  $\times 1$  &  $\times 10$ & $\times 50$ \\
\hline
 1.0  &   6 &   2 &   0 \\
 1.25 &  18 &   4 &   0 \\
 1.5  &  37 &  10 &   0 \\
 2.0  &  95 &  32 &   2 \\
 3.0  & 240 & 112 &  12 \\
 5.0  & 367 & 289 &  73 \\
10.0  & 382 & 377 & 274 \\
\hline
\end{tabular}
\end{center}
\caption{Impact of the stellar variability on the
integrated number of detections expected for COROT. At the
timescale of transits, the solar variability as estimated from
SOHO data is $\simeq 50$~ppm rms. The crowding effect
(Sect.~\ref{sub:crowding}) is expected to remove $\approx 10\%$
from the detections quoted here.} \label{tab:var}
\end{table}
\subsection{Background induced variability} \label{sub:crowding}
In most stellar fields close to the Galactic Plane, the number of
background (BG) stars increases as one goes deeper, typically by a
factor 2.3 per magnitude. Consequently, in any target star mask,
there will be some flux contribution by BG stars, some of which
are variable. This will add to the photometric measurement an
extra noise with a time dependence similar to that of the target
star variability. In this section, we derive a rough estimate of
the fraction of the detections that are lost because of this extra
noise.

Variability of dwarf and giant stars has been measured by the
Geneva group and the Hipparcos mission (Grenon \cite{grenon93},
Eyer and Grenon \cite{eyer}).
Crudely, the result can be described as:
\begin{itemize}
\item no detected variability for 90\% of dwarfs; \item 10\% of
dwarfs have $1.4\,10^{-2}$ variability rms; \item 98\% of giants
have, at most, $2\,10^{-3}$ variability rms; \item 2\% of giants
have, at most, $5\,10^{-2}$ variability rms.
\end{itemize}
These variabilities have been measured on timescales of one to few
months whereas transit search by COROT is mostly concerned with
timescales of 15 hours or less. In the absence of information at
proper timescales, the latter is derived from the former using the
frequency power spectral density of the solar variability
(DIARAD/SOHO, Batalha \cite{batalha}). Fortunately, this scaling
is not very sensitive to the exact frequencies selected.
Typically, a 0.16 scaling factor is found from longer (months) to
shorter (hours) timescales.

\begin{figure}
\centering
\includegraphics[width=8cm]{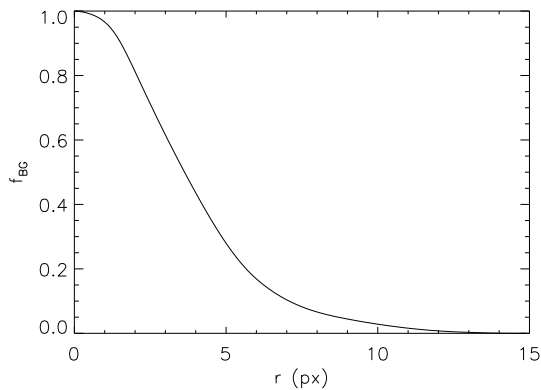}
\caption{Fraction of a background star flux included in a target
mask vs. the distance between their photocenters, once averaged
over all relative orientations.} \label{fig:f_bg}
\end{figure}

The fraction $f_\mathrm{BG}$ of the BG star flux included in the
target mask depends on the relative orientations of the target mask
and the BG star PSF, and on the distance $r$ between their
photocenters. Once averaged over all relative orientations,
$f_\mathrm{BG}$ is a decreasing function with half maximum at
${r=3.6}$~px for a 85~px mask (Fig.~\ref{fig:f_bg}).
A Monte-Carlo simulation is performed for every bin of the target
star magnitude $V$, using the BG stellar densities provided by the
Besan\c con model and the above variability laws. Thus, we compute
histograms of the number of target stars affected by variable BG
stars vs. the standard deviation $\sigma_\mathrm{BG}$ of this extra
variability noise. This effect is considered as perturbing
significantly the detection if it increases the standard deviation
of the total noise by 20\%, i.e. ${\sigma'_n = 1.2 \, \sigma_n}$
or ${\sigma_\mathrm{BG} = 0.66 \, \sigma_n}$ as all sources of noise are
uncorrelated. For instance, the extra variability would prevent
the detection of a planet with a radius $R_\mathrm{P} = 1.5 \,
\mathrm{R}_\oplus$ orbiting a K2 star at $a_r=0.05$~AU if
${\sigma_\mathrm{BG} > 1.02 \, (R_\mathrm{P}/R_\star)^2 \, N_e}$
(Eq.~(\ref{eq:Rp})).

For every bin of magnitude, the number of target stars for which
BG stars induce such a level of extra noise is deduced from the
histograms of the Monte-Carlo simulation. The fraction of lost
detections is obtained by dividing the number of polluted targets
by their total number. Then a weighted average is calculated over
the different bins, with the expected number of detections as a
weight. The loss amounts to 13\% for ${R_\mathrm{P} = 1.5 \,
\mathrm{R}_\oplus}$ and 10\% for ${R_\mathrm{P} = 3 \,
\mathrm{R}_\oplus}$. This fraction seems to vary slowly with the
mask size: for a smaller mask with 60~px, the lost becomes
respectively 11\% and 8.5\%. For mask sizes to be used by COROT,
it is approximatively proportional to the square root of the
number of pixels in the mask.

As a conclusion, \emph{the extra variability induced by BG stars
is a systematic effect that causes a loss of ${\approx 10\%}$ of
the detections}. This impact, although limited, is not negligible.

\subsection{Background eclipsing binaries} \label{sub:binaries}
Most of the time, eclipsing binaries are readily distinguished from
transiting planets by the fact that they lead to two transits of
unequal depth, both lasting longer than expected for a planet. Low
amplitude grazing transits of (almost) identical stellar
companions may be distinguished from a unique transiting planet
with half the revolution period thanks to the characteristic V
shape of grazing transits as opposed to a nearly flat shape for
planets (e.g. Borucki et al. \cite{borucki01}). Low S/N candidates
for which this criterion could not be applied would be the subject
of complementary spectroscopic observations for a definitive
classification.

A tricky case happens when a partly overlapping BG star
(but not the foreground one) is an eclipsing binary of identical
components or features only primary transits because the companion
is too faint. It may then mimic a planetary transit around the
foreground star. If the magnitude of the foreground star is $V \le
14$, then a color discrimination could be done: only if the
superposition of both PSFs is almost perfect, spurious transits
due to the BG star would not be present in all color channels.
Cases where $V > 14$ are more difficult. At the beginning of every
150 day observation of a given field, it is planned to make a long
exposure (typically a couple of hours) to select the target stars
and to define their corresponding aperture masks. This exposure
will allow us to detect potential polluting BG stars. For every
planet candidate orbiting a star potentially polluted,
complementary spectroscopic observations would have to be
conducted as well.
%
%________________________________________________ Section 6 / Conclusion
%
\section{Conclusion}
With these simulations, we have shown that COROT has the capacity
to detect numerous exoplanets, not only Jupiter and Uranus-class
ones, but also hot terrestrial planets, if they exist. Hot
terrestrial planets should be mainly gathered around 14-15th
magnitude K2--M2 stars, Uranus-class planets around 15th magnitude
G2 stars and giant ones around 15--16th magnitude F7--G2 stars.
The number of detections increases with the magnitude of the
parent star up to $V=16$, thus reflecting the original stellar
distribution. However, only hot and frequently transiting planets
may be found around faint stars.

Besides, we have evaluated the impact of a high stellar
variability noise and of the crowding effect. If all star
variabilities were to be orders of magnitude above the Sun level
(an unlikely situation), no terrestrial planet could be within reach
(apart maybe from the use of the color information). Finally,
detection estimates must take into account the crowding effect
that may cause a loss of about 10\% of the detections by inducing
an extra variability noise.

As a final remark, let us point out that COROT will bring the
first data about the abundances and the orbital distributions of
Uranus-class to terrestrial exoplanets. These constitute key information
for the ambitious followers like Darwin (L\'eger et al.
\cite{leger}, Fridlund et al. \cite{fridlund}) and NASA's
Terrestrial Planet Finder (Beichmann et al. \cite{beichmann}).

%
%______________________________________________________ Acknowledgements
%
\begin{acknowledgements}
We are very grateful to Dr.~Michel Grenon for valuable discussions
on stellar variability, and to Dr.~Marc Ollivier for correcting a
few inaccuracies and for his help on taking into account the
jitter noise.
\end{acknowledgements}
%
%____________________________________________________________ References
%

%
\end{document}